\documentclass[aps,prl,showpacs,
amsmath,amssymb,floatfix,nofootinbib]{revtex4}
\usepackage{graphicx} \usepackage{dcolumn} \usepackage{bm}
\usepackage{tipa} \usepackage{CJK}
\newcommand{\comment}[1]{}
\newcommand{\lp}{\underline{p}}
\newcommand{\lo}{\underline{\omega}}
\newcommand{\up}{\overline{p}}
\newcommand{\uo}{\overline{\omega}}

\newcommand{\E}{{\cal E}}
\newcommand{\FM}{{\mathbb Q}'}
\newcommand{\FN}{{\mathbb P}'}
\renewcommand{\P}{\hat{P}}
\newcommand{\Q}{\hat{Q}}
\newcommand{\p}{{P}}
\newcommand{\q}{{Q}}
\renewcommand{\k}{{K}}
\newcommand{\g}{g}
\renewcommand{\I}{I}
\renewcommand{\i}{I}
\newcommand{\0}{{0}}
\begin{document}
\draft 
\title{Imprecise probability for non-commuting observables}

\author{Armen E. Allahverdyan}

\address{
Yerevan Physics Institute, Alikhanian Brothers Street 2,
Yerevan 375036,  Armenia}

\begin{abstract}It is known that non-commuting observables in quantum
  mechanics do not have joint probability. This statement refers to
  the precise (additive) probability model. I show that the joint
  distribution of any non-commuting pair of variables can be
  quantified via upper and lower probabilities, i.e. the joint
  probability is described by an interval instead of a number
  (imprecise probability). I propose transparent axioms from which the
  upper and lower probability operators follow. They depend only on
  the non-commuting observables and revert to the usual expression for
  the commuting case.
\end{abstract}

\pacs{03.65.-w, 03.67.-a}

\maketitle

%\pacs{03.65.-w}{Quantum mechanics}
%\pacs{03.67.-a}{Quantum information}

Non-commuting observables in quantum mechanics do not have a joint
probability \cite{wigner_no_go,hudson_soto,gudder,deMuynck,Busch} [see
section 1.1 of the Supplementary Material for a reminder].  This is
the departure point of quantum mechanics from classical probabilistic
theories \cite{malley}; it lies in the core of all quantum oddities.
There are various quasi-probabilities (e.g., Wigner function) which
have features of joint probability for (loosely defined) semiclassical
states \cite{hillery,ferrie,ballentine}. Quasi-probabilities do have
two problems: {\it (i)} they (must) get negative for a class of
quantum states, thereby preventing {\it any} probabilistic
interpretation for them \footnote{Negative probabilities were not
  found to admit a direct physical meaning \cite{mueck} (what can be
  less possible, then the impossible?).  In certain cases what seemed
  to be a negative probability was later on found to be a local value
  of a physical quantity, i.e. physically meaningful, but not a
  probability \cite{mueck}. Mathematical meaning of negative
  probability is discussed in Refs.~\cite{khren,gabor}.}. {\it (ii)}
Even if the quasi-probability is positive on a certain state, it is
not unique, i.e. there can be other (equally legitimate)
quasi-probability that is positive (and has other expected features of
probability) on this state, e.g. in quantum optics there is Wigner
function, P-function, Terletsky-Margenau-Hill function {\it
  etc}. Despite of the drawbacks, quasi-probabilities do have many
applications \cite{hillery,ferrie,gardiner,blako,armen}, since they
still possess certain features of joint probability, e.g. they
reproduce the marginals
\cite{hudson_soto,deMuynck,hillery,ferrie,armen}. One can relax this
requirement \footnote{Employing instead the unbiasedness: the averages
  of the non-commuting quantities are reproduced correctly
  \cite{arthurs}.}, as done for joint measurements of non-commuting
variables \cite{deMuynck,Busch,de,arthurs,abn,yuko}. Such measurements
have to be approximate, since they operate on an arbitrary initial
state \cite{deMuynck,Busch}. They produce positive probabilities for
the measurement results, but it is not clear to which extent these
probabilities are intrinsic \cite{uffink}, i.e. to which extent they
characterize the system itself, and not approximate measurements
employed. Alternatively, one can consider two consecutive measurements
of the non-commuting observables \cite{bub,hughes}. These two-time
probabilities do not (generally) qualify for the joint probability of
the non-commuting observables; see section 1.2 of the Supplementary
Material.

It is assumed that the sought joint probability is linear over the
state (density matrix). If this condition is skipped, there are
positive probabilities that correctly reproduce marginals for
non-commuting observables \cite{cohen_zap,yuko}, e.g. simply the
product of two marginals \cite{ballentine}. However, they do not
reduce to the usual form of the joint quantum probability for {\it
  commuting} observables \footnote{Given two projectors $\p$ and $\q$
  and state $\rho$, this product is ${\rm tr}(\rho\p)\,{\rm
    tr}(\rho\q)$, while the correct form for $\p\q=\q\p$ is ${\rm
    tr}(\rho\p\q)$.}; hence their physical meaning is unclear
\cite{ballentine}.

The statement on the non-existence of joint probability concern the
usual {\it precise and additive} probability. This is not the only
model of uncertainty. It was recognized since early days of
probability theory \cite{shafer} that the probability need not be
precise: instead of being a definite number, it can be a definite
interval \cite{good,kuz,walley,fine_freq}; see \cite{bumbarash} for an
elementary introduction.

Instead of a precise probability for an event $E$, the measure of
uncertainty is now an interval $[\lp(E),\up(E)]$, where
$0\leq\lp(E)\leq \up(E)$ are called lower and upper probabilities,
respectively. Qualitatively, $\lp(E)$ ($1-\up(E)$) is a measure of a
sure evidence in favor (against) of $E$. The event $E$ is surely more
probable than $E'$, if $\lp(E)\geq \up(E')$. The usual probability is
recovered for $\lp(E)=\up(E)$. Two different pairs $[\lp(E),\up(E)]$
and $[\lp'(E),\up'(E)]$ can hold simultaneously (i.e they are
consistent), provided that $\lp'(E)\leq \lp(E)$ and $\up'(E)\geq
\up(E)$ for all $E$. In particular, every imprecise probability
is consistent with $\lp'(E)=0$, $\up'(E)=1$. 

% The latter ``nothing is known'' situation cannot be represented by
% usual probabilities, the simplest example showing that imprecise
% probabilities can model types of uncertainty that are not captured by
% the precise model.

It is not assumed that for all $E$ there is a true (precise, but
unknown) probability that lies in $[\lp(E),\up(E)]$. This assumption
is frequently (but not always \cite{fine_1994}) made in applications
\cite{kuz,walley}, and it did motivate the generalized Kolmogorovian
axiomatics of imprecise probability \cite{fine_freq}; see section 2.1
of the Supplementary Material. Imprecise joint probabilities in
quantum mechanics are to be regarded as fundamental entities, not
reducible to a lack of knowledge. They do need an independent
axiomatic ground.

My purpose here is to propose a transparent set of conditions (axioms)
that lead to quantum lower and upper joint probabilities. They depend
only on the involved non-commuting observables (and on the quantum
state).

% The assumption is legitimate in statistics, where one bounds the
% unknown (additive) probability via a finite number of observations
% \cite{kuz}. It is not forbidden in subjective theories, where one aims
% at quantifying an uncertain human opinion via probabilities
% \cite{walley}.

{\it Previous work.} In 1967 Prugovecki tried to describe the joint
probability of two non-commuting observables in a way that resembles
imprecise probabilities \cite{prug}. But his expression was not
correct (it still can be negative) \cite{ballentine}; see also
\cite{khren} in this context. In 1991 Suppes and Zanotti proposed a
local upper probability model for the standard setup of Bell
inequalities (two entangled spins) \cite{suppes_zanotti}; see also
\cite{barros,hartmann}. The formulation was given in the classical
event space of hidden variables, and it is not unique even for the
particular case considered. It violates classical observability
conditions for the imprecise probability
\cite{fine_1994,fine_freq,suppes_zanotti}. In particular, no lower
probability exists in this scheme. Despite of such drawbacks, the
pertinent message of \cite{suppes_zanotti} is that one should attempt
at quantum applications of the upper probabilities that go beyond its
classical axioms. More recently, Galvan attempted to empoy (classical)
imprecise probabilities for describing quantum dynamics in
configuration space \cite{galvan}. For a general discussion on quantum
versus classical probabilities see \cite{khren_book}.

% In \cite{fine,ochs} imprecise probabilities in quantum mechanics were
% mentioned in the context of comparative (modal) probability.

{\it Notations.} All operators (matrices) live in a finite-dimensional
Hilbert space $\mathbb{H}$. For two hermitean operators $Y$ and $Z$,
$Y\geq Z$ (larger or equal) means that all eigenvalues of $Y-Z$ are
non-negative, i.e. $\langle \psi|(Y-Z)\psi\rangle \geq 0$ for any
$|\psi\rangle \in \mathbb{H}$.  The direct sum $Y\oplus Z$ of two
operators refers to the block-diagonal matrix: $Y\oplus Z=\left(
     \begin{array}{cc}
Y & 0 \\
0 & Z \\
\end{array}\right)$. The range ${\rm ran}(Y)$ of $Y$ is the
subspace of vectors $Y|\psi\rangle$, where
$|\psi\rangle\in\mathbb{H}$. For orthogonal (sub)spaces $\mathbb{A}$
and $\mathbb{B}$, the space $\mathbb{A}\oplus\mathbb{B}$ is formed by
all vectors $|a\rangle+|b\rangle$, where $|a\rangle\in \mathbb{A}$ and
$|b\rangle\in \mathbb{B}$. $\I$ is the unity operator of
$\mathbb{H}$. $\i_{n}$ and $\0_{n}$ are the $n\times n$ unity and zero
matrices, respectively.

{\it Axioms for quantum imprecise probability}. Existing axioms for
imprecise probability are formulated on a classical event space with
usual notions of con- and disjunction and complemention 
\cite{good,kuz,walley,fine_freq,good}; see section 2 of the Supplementary
Material for a reminder. For quantum probability it is natural to
start from a Hilbert space and introduce upper and lower probabilities
as operators. The axioms below require only the most basic feature of
upper and lower probability and demand its consistency with the
quantum joint probability whenever the latter is well-defined.

The usual quantum probability can be defined over (hermitean)
projectors $\p=\p^2$ \cite{jauch,jauch_book}. A projector generalizes
the classical notion of characteristic function. Each $\p$ uniquely
relates to its eigenspace ${\rm ran}(\p)$. $\p$ refers to a set of
hermitean operators $\{{\cal P}\}$: 
\begin{eqnarray}
  \label{eq:81}
[{\cal P},\p]\equiv {\cal
  P}\p-\p{\cal P}=0. 
\end{eqnarray}
$\p$ is a projector to an eigenspace of ${\cal
  P}$ or to a direct sum of such eigenspaces, i.e. $\p$ refers to an
eigenvalue of ${\cal P}$ or to a union of several eigenvalues.  The
quantum (precise and additive) probability to observe $\p=1$ is ${\rm
  tr}[\rho \p]$, where the density matrix $0\leq \rho\leq \I$ defines
the quantum state \cite{deMuynck,Busch,jauch,jauch_book}.

\comment{
Let we are given two sets of non-commuting, $[P_k,Q_i]\not =0$,
projectors [$\delta_{ij}$ is the Kronecker symbol]:
\begin{eqnarray}
  \label{eq:19}
&&  \sum_{k=1}^{n_P} P_k=\I, ~~~ P_kP_{k'}=\delta_{kk'}P_k, ~~ n_P\leq {\rm
dim}\mathbb{H}, \\
&&  \sum_{i=1}^{n_Q} Q_i=\I, ~~~ Q_iQ_{i'}=\delta_{ii'}Q_i, ~~ n_Q\leq {\rm
dim}\mathbb{H}.
  \label{eq:191}
\end{eqnarray}
Eq.~(\ref{eq:19}) refers to a class of hermitean operators $\{{\cal
  P}\}$ so that $[{\cal P},P_k]=0$ for $1\leq k\leq n_p$. Each
$P_k$ is a projector to an eigenspace of ${\cal P}$ or to a direct sum
of such eigenspaces, i.e. each $P_k$ refers to an eigenvalue of
${\cal P}$ or to a union of several eigenvalues. Eq.~(\ref{eq:191})
admits a similar interpretation.  Eq.~(\ref{eq:19}, \ref{eq:191})
imply $0\leq P_k,Q_i\leq \I$ for all $k$ and $i$.
}

Let $\q$ be another projector which refers to the set $\{{\cal Q}\}$ of
observables. Generally, $[\p,\q]\not=0$.  Given the density matrix
$\rho$, we seek upper and lower joint probabilities of $\p$ and $\q$
(i.e. of the corresponding eigenvalues of ${\cal P}$ and ${\cal Q}$):
\begin{eqnarray}
  \label{ax0}
  \up(\rho; \p, \q)={\rm tr}(\rho \,\uo(\p, \q)\,), ~~~   
\lp(\rho; \p, \q)={\rm tr}(\rho\, \lo(\p, \q)\,),
\end{eqnarray}
where $\lo(\p, \q)$ and $\uo(\p, \q)$ are hermitean operators. Their
dependence on $\p$ and $\q$ can be expressed via Taylor series. 
We impose the following conditions (axioms):
\begin{eqnarray}
  \label{ax1}
0\leq  \lo(\p, \q)\leq   \uo(\p, \q)\leq \I,
\end{eqnarray}
\begin{eqnarray}
  \label{ax2}
 \lo(\p, \q)=  \lo(\q,\p), ~~~ 
  \uo(\p, \q)=  \uo(\q,\p),
\end{eqnarray}
\begin{eqnarray}
  \label{ax3}
\lo(\p, \q)= \uo(\p, \q)=\p \q, ~~ {\rm if}~~ [\p, \q]=0,
\end{eqnarray}
\begin{eqnarray}
  \label{ax4}
{\rm tr}(\rho \,\lo(\p, \q)\,)\leq {\rm tr}(\rho \,\p
\q\,)\leq {\rm tr}(\rho\, \uo(\p, \q)\,),~~ {\rm if}~~ [\p, \rho]=0,
~~ {\rm or ~ if}~~ [\rho, \q]=0. 
\end{eqnarray}
\begin{eqnarray}
  \label{ax5}
  [\omega(\p, \q),\q]=[\omega(\p, \q),\p]=0, ~~ \omega=\lo,\uo.   
\end{eqnarray}
Eq.~(\ref{ax0}) implies that $\lp$ and $\up$ depend on $\{{\cal P}\}$
and $\{{\cal Q}\}$ only through $\p$ and $\q$. This non-contextuality
feature holds also for the ordinary (one-variable) quantum probability
\cite{bell,smerd}. Provided that the operators $\lo$ and $\uo$ are
found, $\lp$ and $\up$ can be found in the usual way of quantum
averages.

Conditions (\ref{ax1}) stem from $0\leq\up (\rho; \p, \q)\leq \lp
(\rho; \p, \q)\leq 1$ that are demanded for all density matrices
$\rho$. Eq.~(\ref{ax2}) is the symmetry condition necessary for the
joint probability. Eq.~(\ref{ax3}) is reversion to the commuting
case. In particular, (\ref{ax3}) ensures $\lo(\p, 0)= \uo(\p, 0)=0$
and $\lo(\p, \I)= \uo(\p, \I)=\p$. Since $\q=\I$ means that $\q$ is
anywhere, the latter equality is the reproduction of the marginal
probability (which cannot be recovered by summation, since the
probability model is not additive). 

For $[\p,\q]=0$ the joint probability is ${\rm tr}(\rho \q \,\p
\,)={\rm tr}(\rho \p \q \,)$.  This expression is well-defined
(i.e. positive, symmetric and additive) also for $[\rho, \q]=0$ or
$[\rho, \p]=0$ (but not necessarily $[\p,\q]=0$). If $[\rho, \q]=0$,
one obtains ${\rm tr}(\rho \q \,\p)$ by measuring $\q$ ($\rho$ is not
disturbed) and then $\p$. Alternatively, one can obtain it by
measuring the average of an hermitean observable
$\frac{1}{2}(\p\q+\q\p)$.  Thus (\ref{ax3}, \ref{ax4}) demands that
$\up(\rho; \p, \q)$ and $\lp(\rho; \p, \q)$ are consistent with the
joint probability ${\rm tr}(\rho \,\p \q\,)$, whenever the latter is
well-defined.

Finally, (\ref{ax5}) means that $\omega(\p, \q)$ ($\omega=\lo,\uo$)
can be measured {\it simultaneously and precisely} with $\p$ or with
$\q$ (on any quantum state), a natural condition for the joint
probability (operators).

If there are several candidates satisfying (\ref{ax1}--\ref{ax5}) we
shall naturally select the ones providing the largest lower
probability and the smallest upper probability. 

{\it CS-representation} will be our main tool. Given the projectors
$\p$ and $\q$, Hilbert space $\mathbb{H}$ can be represented as a
direct sum \cite{dix,halmos,hardegree} [see also section 3 of the
Supplementary Material]
\begin{eqnarray}
  \label{eq:36}
\mathbb{H}=\mathbb{H}'\oplus \mathbb{H}_{11}
\oplus \mathbb{H}_{10}\oplus \mathbb{H}_{01}\oplus
\mathbb{H}_{00},
\end{eqnarray}
where the sub-space $\mathbb{H}_{\alpha\beta}$ of dimension
$m_{\alpha\beta}$ is formed by common eigenvectors of $\p$ and $\q$
having eigenvalue $\alpha$ (for $\p$) and $\beta$ (for
$\q$). Depending on $\p$ and $\q$ every sub-space can be absent; all
of them can be present only for ${\rm dim}\mathbb{H}\geq 6$. Now
$\mathbb{H}_{11}={\rm ran}(\p)\cap{\rm ran}(\q)$ is the intersection
of the ranges of $\p$ and $\q$. $\mathbb{H}'$ has even dimension $2m$
\cite{halmos,hardegree}, this is the only sub-space in (\ref{eq:36})
that is not formed by common eigenvectors of $\p$ and $\q$. There
exists a unitary transformation
\begin{eqnarray}
  \label{eq:80}
\p=U\P U^\dagger, ~~~ \q=U\Q U^\dagger, ~~~ UU^\dagger=\I,   
\end{eqnarray}
so that $\P$ and $\Q$ get the following
block-diagonal form related to (\ref{eq:36}) \cite{halmos}:
\begin{eqnarray}
  \label{eq:cs}
  \Q=Q'\oplus \i_{m_{11}} \oplus \i_{m_{10}} \oplus \0_{m_{01}}\oplus
  \0_{m_{00}}, ~~ Q'\equiv\left(
     \begin{array}{cc}
\i_m & \0_m \\
\0_m & \0_m \\
\end{array}\right),  \\
  \P=P'\oplus \i_{m_{11}} \oplus \0_{m_{10}} \oplus \i_{m_{01}} \oplus \0_{m_{00}}, ~~
P'\equiv\left(
     \begin{array}{cc}
C^2 & CS \\
CS & S^2 \\
\end{array}\right), 
  \label{eq:cs1}
\end{eqnarray}
where $C$ and $S$ are invertible square matrices of the same size
holding
\begin{eqnarray}
  \label{eq:35}
C^2+S^2=\i_m, ~~ [C,S]=0. 
\end{eqnarray}
Now ${\rm ran}(P')$ and ${\rm ran}(Q')$ are sub-spaces of
$\mathbb{H}'$. One has $C=\cos T$ and $S=\sin T$, where $T$ is the
operator analogue of the angle between two spaces.
$\mathbb{H}_{m_{\alpha\beta}}$ are absent, if $P$ and $Q$ do not have
any common eigenvector. This, in particular, happens in
${\rm dim}(\mathbb{H})=2$.

{\it The main result}. Note that if (\ref{ax1}--\ref{ax5}) holds for
$\p$ and $\q$, they hold as well for $\P$ and $\Q$, because
$\omega(\p,\q)=U^\dagger\omega(\P,\Q)U$ for $\omega=\lo,\uo$. 
Section 4 of the Supplementary Material shows how to get 
$\uo(\P,\Q)$ and $\lo(\P,\Q)$ from (\ref{ax1}--\ref{ax5}) 
and (\ref{eq:cs}, \ref{eq:cs1}): 
\begin{eqnarray}
  \label{eq:37}
&&\uo(\P,\Q)= 
\left(
     \begin{array}{cc}
C^2 & \0_m \\
\0_m & C^2 \\
\end{array}\right)\oplus \i_{m_{11}}\oplus \0_{m_{10}+m_{01}+m_{00}}, \\
&&\lo(\P,\Q)= 
\0_{2m}\oplus \i_{m_{11}}\oplus \0_{m_{10}+m_{01}+m_{00}}.
  \label{eq:3737}
\end{eqnarray}
Let $\g(P,Q)=\g(Q,P)$ be the projector onto intersection ${\rm
  ran}(P)\cap{\rm ran}(Q)$ of ${\rm ran}(P)$ and ${\rm ran}(Q)$. We
now return from (\ref{eq:37}, \ref{eq:3737}, \ref{eq:80}) to original
projectors $\p$ and $\q$ [see section 4 of the Supplementary Material]
and obtain the main formulas:
\begin{eqnarray}
  \label{eq:33}
&&  \lo(\p,\q)=  \g(\p,\q),\\
&& \uo (\p,\q)= \I-(\p-\q)^2-\g(\I-\p,\I-\q). 
  \label{eq:333}
\end{eqnarray}
For $[\p,\q]=0$, $\g(\p,\q)=\p\q$, and we revert to $\lo(\p,\q)=
\uo(\p,\q)=\p\q$. Note that
$[\p,(\p-\q)^2]=[\q,(\p-\q)^2]=0$. 

{\it Physical meaning of $\lo$ and $\uo$.} When looking for a joint
probability defined over two projectors $\p$ and $\q$ one wonders
whether it is just not some (operator) mean of $\p$ and $\q$. For
ordinary numbers $a\geq 0$ and $b\geq 0$ there are 3 means: arithmetic
$\frac{a+b}{2}$, geometric $\sqrt{ab}$ and harmonic
$\frac{2ab}{a+b}$. Now (\ref{eq:33}) is precisely the operator
harmonic mean of $\p$ and $\q$ \cite{anderson_duffin1}
\begin{eqnarray}
  \label{eq:77}
  \g(\p,\q)=2\p(\p+\q)^-\,\q=2\q(\p+\q)^-\,\p,
\end{eqnarray}
where $A^-$ is the inverse of $A$ if it exists, otherwise it is the
pseudo-inverse; see section 5 of the Supplementary Material for
various representations of $\lo(\p,\q)$ and $\uo(\p,\q)$. More
familiar formula is 
\begin{eqnarray}
  \label{eq:20}
  \g(\p,\q)={\rm lim}_{n\to\infty}\q(\p\q)^n={\rm lim}_{n\to\infty}\p(\q\p)^n.
\end{eqnarray}

The intersection projector $\g(\p,\q)$ appears in
\cite{jauch,jauch_book,jauch_piron,busch,bell}. It was stressed that
$\g(\p,\q)$ cannot be a joint probability for non-commutative $\p$ and
$\q$ \cite{de}. Its meaning is clear by now: it is the lower
probability for $\p$ and $\q$.  $\g(\p,\q)$ is non-zero only if ${\rm
  tr}(\p)\geq 2$ (or ${\rm tr}(\q)\geq 2$), since two different rays
cross only at zero.

Let us now turn to $\uo$. The transition probability between 2 pure
states is determined by the squared cosine of the angle between them:
$|\langle\psi|\phi\rangle|^2=\cos^2\theta_{\phi\psi}$.
Eq.~(\ref{eq:37}) shows that $\uo(\p,\q)$ depends on $C^2=\cos^2 T$,
where $T$ is the operator angle between $\P$ and $\Q$. Note from
(\ref{eq:cs}, \ref{eq:cs1}) that the eigenvalues $\lambda$ of $\p\q$,
which hold $0<\lambda<1$ are the eigenvalues of $C^2$, and|as seen
from (\ref{eq:37})|they are also (doubly-degenerate) eigenvalues of
$\uo(\p,\q)$. Thus we have a physical interpretation not only for
${\rm tr}(\p\q)$ (transition probability), but also for eigenvalues of
$\p\q$ ($\p\q$ and $\q\p$ have the same eigenvalues).

Eqs.~(\ref{eq:37}, \ref{eq:3737}, \ref{eq:80}) imply that the upper
and lower probability operators can be measured simultaneously on any
state [cf. (\ref{ax5})]:
\begin{eqnarray}
  \label{doris}
[\lo(\p,\q),\uo(\p,\q)]=0.   
\end{eqnarray}
The operator $\uo(\p,\q)-\lo(\p,\q)$ quantifies the uncertainty for
joint probability, the physical meaning of this characteristics of
non-commutativity is new. 

% As seen from (\ref{eq:36}--\ref{eq:cs1}), $\I-\g(\I-\p,\I-\q)$ in
% (\ref{eq:333}) is the projector of ${\rm ran}(\p)+{\rm ran}(\q)$ (the
% linear space formed by superpositions of all vectors from ${\rm
%   ran}(\p)$ and form ${\rm ran}(\q)$), i.e. it has a geometric
% meaning.

Section 7 of the Supplementary Material calculates the upper and lower
probabilities for several examples. 

Note that the conditional (upper
and lower) probabilities are straighforward to define,
e.g. [cf. (\ref{ax0})]: $\up(\rho; \p| \q)=\up(\rho; \p, \q)/{\rm
  tr}(\rho \q)$. 

The distance between two probability intervals
$[\lp,\up]$ and $[\lp',\up']$ can be calculated via the Haussdorff
metric \cite{alefeld} 
\begin{eqnarray}
  \label{eq:76}
{\rm max}\left[\, |\lp-\lp'|, \, |\up-\up'| \,
\right],   
\end{eqnarray}
which nullifies if and only if $\lp=\lp'$ and $\up=\up'$, and which
reduces to the ordinary distance $|p-p'|$ for usual (precise)
probabilities.
Now
\begin{eqnarray}
  \label{eq:32}
  {\rm tr}(\rho\, \lo(\p_1,\q_1))>{\rm
    tr}(\rho\, \uo(\p,\q)),  
\end{eqnarray}
means that the pair of projectors $(\p_1,\q_1)$ is surely more
probable (on $\rho$) than $(\p,\q)$; see section 7 of the
Supplementary Material for examples. Note from (\ref{eq:33},
\ref{eq:333}) that if
\begin{eqnarray}
  \label{eq:61}
{\rm tr}(\rho\,\omega(\p,\q))>{\rm tr}(\rho\,\omega(\I-\p,\I-\q )),  
\end{eqnarray}
holds for $\omega=\lo$, then it also hods for $\omega=\uo$ (and {\it
  vice versa}). Though in a weaker sense than (\ref{eq:32}),
(\ref{eq:61}) means that $\p$ and $\q$ together is more probable than
neither of them together (which is the pair $(\I-\p,\I-\q)$).
Eqs.~(\ref{eq:32}, \ref{eq:61}) are examples of comparative (modal)
probability statements; see \cite{ochs} in this context. 

Further features of $\lo$ and $\uo$ are uncovered when looking at a
monotonic change of their arguments; see section 6 of the
Supplementary Material. Section 7 discusses concrete examples.

{\it Summary}. My main message is that while joint precise probability
for non-commuting observables does not exist, there are well-defined
expressions for upper and lower imprecise probabilities. They can have
immediate applications as shown in section 7 of the Supplementary
Material for simple examples. Not less important are the open question
suggested by this research, e.g. what is the most convenient way of
defining averages with respect to quantum imprecise probability, or
are there even more general axioms that involve the density matrix
non-lineary and reduce to the linear situation when (\ref{ax3},
\ref{ax4}) (effective commutativity) holds.

I thank K.V. Hovhannisyan for discussions.

\clearpage

\section{{\large Supplementary Material} }
\begin{center}
{\bf Imprecise probability for non-commuting observables}
by Armen E. Allahverdyan
\end{center}

This Supplementary Material consists of seven sections. All of them
can be read independently from each other. 

Sections 1, 2 and 3 recall, respectively, the no-go statements for
the joint quantum probability, generalized axiomatics for the
imprecise probability and the CS-representation. This material is not
new, but is presented in a focused form, adapted from several
different sources.

Section 4 contains the derivation of the main result, while sections 5
and 6 demonstrate various feature of quantum imprecise probability.

Section 7 illustrates it with simple physical examples.

\section{0. Notations}

We first of all recall the employed notations. All operators
(matrices) live in a finite-dimensional Hilbert space
$\mathbb{H}$. For two hermitean operators $Y$ and $Z$, $Y\geq Z$
means that all eigenvalues of $Y-Z$ are non-negative, i.e. $\langle
\psi|(Y-Z)\psi\rangle \geq 0$ for any $|\psi\rangle \in \mathbb{H}$.
The direct sum $Y\oplus Z$ of two operators refers to the following
block-diagonal matrix: $$Y\oplus Z=\left(
     \begin{array}{cc}
Y & 0 \\
0 & Z \\
\end{array}\right).$$ 
${\rm ran}(Y)$ is the range of $Y$ (set
of vectors $Y|\psi\rangle$, where $|\psi\rangle\in\mathbb{H}$). $\I$
is the unity operator of $\mathbb{H}$. ${\rm ker}(Y)$ is the
subspace of vectors $|\phi\rangle$ with $Y|\phi\rangle=0$.

$\i_{n}$ and $\0_{n}$ are the $n\times n$ unity and zero matrices,
respectively.

In the direct sum of two sub-spaces, $\mathbb{H}\oplus\mathbb{G}$ it
is always understood that $\mathbb{H}$ and $\mathbb{G}$ are
orthogonal. The vector sum of (not necessarily orthogonal) sub-spaces
$\mathbb{A}$ and $\mathbb{B}$ will be denoted as
$\mathbb{A}+\mathbb{B}$. This space is formed by all vectors
$|\psi\rangle+|\phi\rangle$, where $|\psi\rangle\in \mathbb{A}$ and
$|\phi\rangle\in \mathbb{B}$.

\section{1. Non-existence of (precise)  joint probability for
  non-commuting observables}

\subsection{1.1 The basic argument}

Given two sets of non-commuting hermitean projectors:
\begin{eqnarray}
  \label{eq:199}
&&  \sum_{k=1}^{n_P} P_k=\I, ~~~ P_kP_i=\delta_{ik}P_k, ~~ n_P\leq n, \\
&&  \sum_{k=1}^{n_Q} Q_k=\I, ~~~ Q_kQ_i=\delta_{ik}Q_k, ~~ n_Q\leq n,
  \label{eq:1999}
\end{eqnarray}
we are looking for non-negative operators $\Pi_{ik}\geq 0$ such that
for an arbitrary density matrix $\rho$
\begin{eqnarray}
  \label{eq:27}
\sum_{ik} {\rm tr}(\rho \Pi_{ik})=1, ~~~
\sum_i  {\rm tr}(\rho \Pi_{ik})={\rm tr}(\rho P_{k}), ~~~
\sum_k  {\rm tr}(\rho \Pi_{ik})={\rm tr}(\rho Q_{i}).
\end{eqnarray}
These relations imply
\begin{eqnarray}
  \label{eq:28}
  \sum_{ik} \Pi_{ik}=\I, ~~~ \Pi_{ik}\leq Q_i, ~~~ \Pi_{ik}\leq P_k.
\end{eqnarray}
Now the second (third) relation in (\ref{eq:28}) implies 
${\rm ran}(\Pi_{ik})\subseteq {\rm ran}(Q_i)$
(${\rm ran}(\Pi_{ik})\subseteq {\rm ran}(P_k)$). Hence 
${\rm ran}(\Pi_{ik})\subseteq {\rm ran}(Q_i)\cap {\rm ran}(P_k)$. 

Thus, if ${\rm ran}(Q_i)\cap {\rm ran}(P_k)=0$ (e.g. when $P_k$ and
$Q_i$ are one-dimensional), then $\Pi_{ik}=0$, which means that the
sought joint probability does not exist.

If ${\rm ran}(Q_i)\cap {\rm ran}(P_k)\not =0$, then the largest
$\Pi_{ik}$ that holds the second and third relation in (\ref{eq:28})
is the projection $g(P_k,Q_i)$ on ${\rm ran}(Q_i)\cap {\rm
  ran}(P_k)=0$. However, the first relation in (\ref{eq:28}) is still
impossible to satisfy (for $[P_i,Q_k]\not =0$), as seen from the
superadditivity feature (\ref{eq:41a}):
\begin{eqnarray}
  \label{eq:29}
  \sum_{ik} g(P_i,Q_k) \leq \sum_{k} g(\sum _i P_i,Q_k)
=\sum_{k} g(\I,Q_k)=\sum_k Q_k=\I. 
\end{eqnarray}

\subsection{1.2 Two-time probability (as a candidate for the joint
  probability)}

Given (\ref{eq:199}, \ref{eq:1999}), we can carry out two successive
measurements. First (second) we measure a quantity, whose
eigen-projections are $\{P_k\}$ ($\{Q_i\}$). This results to the
following joint probability for the measurement results [$\rho$ is the
density matrix]
\begin{eqnarray}
  \label{eq:599}
  {\rm tr}(\, Q_iP_k\rho P_k\,).
\end{eqnarray}
Likewise, if we first measure $\{Q_i\}$ and then $\{P_k\}$, we obtain
a quantity that generally differs from (\ref{eq:599}):
\begin{eqnarray}
  \label{eq:699}
  {\rm tr}(\,P_k Q_i\rho Q_i \,).
\end{eqnarray}
If we attempt to consider (\ref{eq:699}) [or (\ref{eq:599})] as a
joint additive probability for $P_i$ and $Q_k$, we note that
(\ref{eq:699}) [and likewise (\ref{eq:599})] reproduces correctly only
one marginal:
\begin{eqnarray}
  \label{eq:629}
\sum_{i=1}^{n_Q} {\rm tr}(\, Q_iP_k\rho P_k\,)= 
{\rm tr}(P_k\rho ), ~~~{\rm but}~~~ \sum_{i=1}^{n_P} {\rm tr}(\,
Q_iP_k\rho P_k\,)\not = {\rm tr}(Q_i\rho ).
\end{eqnarray}

One can attempt to interpret the mean of (\ref{eq:599}, (\ref{eq:699}) 
\begin{eqnarray}
  \label{eq:799}
\mu (\rho;P_k,Q_i)=
\frac{1}{2}[  
{\rm tr}(\,P_k Q_i\rho Q_i \,)+  {\rm tr}(\,Q_i P_k\rho P_k \,)]  
={\rm tr}(\,\rho \, \frac{P_k Q_iP_k+  Q_iP_kQ_i}{2}\, ),
\end{eqnarray}
as a non-additive probability. This object is linear over $\rho$,
symmetric (with respect to interchanging $P_k$ and $Q_i$),
non-negative, and reduces to the additive joint probability for
$[P_k,Q_i]=0$. The relation $\mu (\rho;P_k,I)={\rm tr}(\rho P_k)$ can
be interpreted as consistency with the correct marginals (once $\mu
(\rho;P_k,Q_i)$ is regarded as a non-additive probability, there is no
point in insisting that the marginals are obtained in the additive
way).

However, the additive joint probability ${\rm tr}(\rho
P_kQ_i)$ is well-defined also for $[\rho,P_k]=0$ (or for
$[\rho,Q_i]=0$). If $[\rho,P_k]=0$ holds, $\mu (\rho;P_k,Q_i)$ is not
consistent with ${\rm tr}(\rho P_kQ_i)$, i.e. depending on $\rho$,
$P_k$ and $Q_i$ both  
\begin{eqnarray}
  \label{eq:62}
  \mu (\rho;P_k,Q_i) >{\rm tr}(\rho P_kQ_i) ~~~ {\rm and}~~~
  \mu (\rho;P_k,Q_i) <{\rm tr}(\rho P_kQ_i)
\end{eqnarray}
are possible. 

To summarize, the two-time measurement results do not qualify as the
additive joint probability, first because they are not unique (two
different expressions (\ref{eq:599}) and (\ref{eq:699}) are possible),
and second because they do not reproduce the correct marginals. If we
take the mean of two expressions (\ref{eq:599}) and (\ref{eq:699}) and
attempt to interpret it as a non-additive probability, it is not
compatible with the joint probability, whenever the latter is
well-defined.

\section{2. Axioms for classical imprecise probability}

\subsection{2.1 Generalized Kolmogorov's axioms}

% $\supset \subset \supseteq \oplus \forall \in \bot \exists \ast \cup
% \cap \star A^\star \dagger \ddag \amalg A^\amalg$
 
Given the full set of events $\Omega$, $\up(.)$ and $\lp(.)$ defined
over sub-sets $A,B,...$ of $\Omega$ (including the empty set $\{0\}$)
satisfy \cite{kuz,walley,fine_freq}:
\begin{eqnarray}
  \label{eq:1}
&&  \lp(\{0\})=0, \\ 
  \label{eq:2}
&& \up(\Omega)=1, \\
  \label{eq:3}
&& \up(A)=1-\lp(\Omega-{A}), \\
  \label{eq:4}
&& \lp(A\cup B)\geq \lp(A)+\lp(B), ~~{\rm if}~~ A\cap B=\{0\}, \\
  \label{eq:5}
&& \up(A\cup B)\leq \up(A)+\up(B), ~~{\rm if}~~ A\cap B=\{0\},
\end{eqnarray}
where $\Omega-{A}$ includes all elements of $\Omega$ that are not in
$A$, and where $A\cap B$ means intersection of two sets; $A\cap
B=\{0\}$ holds for elementary events.

Here are some direct implications of (\ref{eq:1}--\ref{eq:5}).
\begin{eqnarray}
  \label{eq:6}
A\supseteq B \Longrightarrow  && \up(A)\geq \up (B),\\
                              && \lp(A)\geq \lp (B).
  \label{eq:7}
\end{eqnarray}
Eq.~(\ref{eq:7}) follows directly from (\ref{eq:4}). Eq.~(\ref{eq:6})
follows from (\ref{eq:4}, \ref{eq:3}). Next relation:
\begin{eqnarray}
  \label{eq:8}
\up(A\cup B)  \geq \up(A)+\lp(B)\geq \lp(A\cup B),
~~{\rm if}~~ A\cap B=0,
\end{eqnarray}
which, in particular, implies 
\begin{eqnarray}
  \label{eq:9}
  \up(A)\geq \lp(A).
\end{eqnarray}
To derive (\ref{eq:8}), note that (\ref{eq:4}, \ref{eq:3}) imply 
$\up (\Omega - A- B)\leq \up(\Omega-A)-\lp(B)$ or 
$\up(\Omega-A)\geq \lp(B)+ \up (\Omega-A-B)$, which is the first
inequality in (\ref{eq:8}). The second inequality is derived via 
(\ref{eq:5}, \ref{eq:3}).

The following inequality generalizes the known relation of the
additive probability theory
\begin{eqnarray}
  \label{eq:10}
  \lp (A)+\lp(B)\leq \up (A\cup B)+\lp(A\cap B)
\end{eqnarray}
To prove (\ref{eq:10}), we denote $A'=A-A\cap B$, which means $A'\cap
B=\{0\}$. Now 
\begin{eqnarray}
  \label{eq:11}
  \up (A\cup B)+ \lp (A\cap B)=\up (A'\cup B)+\lp (A\cap B)&&
  \nonumber\\
  \label{eq:12}
\geq && \up (A')+\lp (A\cap B)+\lp (B)\\
\geq && \lp (A)+\lp (B)
  \label{eq:13}
\end{eqnarray}
where in (\ref{eq:12}) [resp. in (\ref{eq:13})]
we applied the first [resp. the second] inequality in (\ref{eq:8}). 

Note that the (non-negative) difference $\Delta p (A)=\up (A)-\lp (A)$
between the upper and lower probabilities also holds the
super-additivity feature (cd. (\ref{eq:5}))
\begin{eqnarray}
  \label{eq:24}
\Delta p(A\cup B)\leq \Delta p(A)+\Delta p(B), ~~{\rm if}~~ A\cap B=0.
\end{eqnarray}

Employing (\ref{eq:8}) one can derive \cite{halpern} for arbitrary
$A_1$ and $A_2$:
\begin{eqnarray}
  \label{gomes1}
 \lp (A_1\cup A_2)+  \lp (A_1\cap A_2)\leq \lp(A_1)+\up(A_2)\leq 
  \up (A_1\cup A_2)+  \up (A_1\cap A_2),      
\end{eqnarray}
\begin{eqnarray}
\lp(A_1)+\lp(A_2)\leq   \lp (A_1\cup A_2)+  \up (A_1\cap A_2)
\leq     \up(A_1)+\up(A_2), 
\end{eqnarray}
\begin{eqnarray}
\lp(A_1)+\lp(A_2)\leq   \up (A_1\cup A_2)+  \lp (A_1\cap A_2)
\leq     \up(A_1)+\up(A_2).
\end{eqnarray}

\subsection{2.2 Joint probability}

The joint probabilities of $A$ and $B$ are now defined as 
\begin{eqnarray}
  \label{eq:59}
  \lp(A,B)=\lp(A\cap B), ~~~   \up(A,B)=\up(A\cap B).
\end{eqnarray}
Employing the distributivity feature
\begin{eqnarray}
  \label{eq:63}
  (A_1\cup A_2)\cap A_3 =   (A_1\cap A_3)\cup (A_2\cap A_3),
\end{eqnarray}
which holds for any triple $A_1,A_2,A_3$, we obtain from
(\ref{eq:4}, \ref{eq:5}) for $B\cap C=\{0\}$
\begin{eqnarray}
  \label{eq:64}
&&  \lp (A,B\cup C)=\lp ((A\cap B)\cup (A\cap C))
\geq   \lp (A,B)+  \lp (A, C), \\
&&  \up (A,B\cup C)=\up ((A\cap B)\cup (A\cap C))
\leq   \up (A,B)+  \up (A, C).
\end{eqnarray}

\subsection{2.3 Dominated upper and lower probability}

The origin of (\ref{eq:1}--\ref{eq:5}) can be related to the {\it
  simplest} scheme of hidden variable(s) \cite{kuz}. One imagines that
there exists a precise probability $P_{\theta}(A)$, where the
parameter $\theta$ is not known. Only the extremal values over the
parameter are known: 
\begin{eqnarray}
  \label{eq:60}
  \up (A)={\rm max}_\theta [P_\theta(A)],~~ \lp
  (A)={\rm min}_\theta [P_\theta(A)],  
\end{eqnarray}
which satisfy (\ref{eq:1}--\ref{eq:5}).

However, it is generally not true that (\ref{eq:1}--\ref{eq:5}) imply
the existence of a precise probability $P_\theta(A)$ that holds
(\ref{eq:60}) \cite{walley}.

\comment{
Somewhat different axioms for the upper and lower probability were
proposed by Good \cite{good}. He also remarks that 
\begin{eqnarray}
  \label{eq:55}
&&  \up (A\cap B)+  \up (A\cup B)\leq   \up (B)+  \up (A), \\
&&  \lp (A\cap B)+  \lp (A\cup B)\geq   \lp (B)+  \lp (A), 
  \label{eq:555}
\end{eqnarray}
where (\ref{eq:55}) and (\ref{eq:555}) imply each other via
(\ref{eq:3}), does not follow from (\ref{eq:1}--\ref{eq:5})
[or equivalent axioms], but nevertheless holds for inner and outer
Lebesgue measures. }

\section{3. Derivation of the CS-representation}

\subsection{3.1 The main theorem}

Let $\FM$ and $\FN$ are two subspaces of Hilbert space $\mathbb{H}'$
that hold ($\FM^\bot$ is the
orthogonal complement of $\FM$)
\begin{eqnarray}
  \label{generic}
  \FM\cap\FN=0, ~~~   \FM\cap\FN^\bot=0, ~~~   \FM^\bot\cap\FN=0,
~~~   \FM^\bot\cap\FN^\bot=0.
\end{eqnarray}
The simplest example realizing (\ref{generic}) is when $\FM$ and $\FN$
are one-dimensional subspaces of a two-dimensional $\mathbb{H}'$. 

Let $\Q'$ and $\P'$ be projectors onto $\FM$ and $\FN$
respectively. Now $\I-\Q'$ is the projector of $\FN^\bot$, and let
$g(\P',\Q')$ be the projector $\FM\cap \FN$. Employing the known
formulas (see e.g. \cite{hardegree}) 
\begin{eqnarray}
  \label{eq:41}
  {\rm tr}(\, \Q-g(\Q,\I-\P)   \,)=  {\rm tr}(\, \P-g(\P,\I-\Q)   \,),
\end{eqnarray}
we get from (\ref{generic})
\begin{eqnarray}
  \label{eq:22}
  {\rm dim}\,\FN = {\rm dim}\,\FM={\rm dim}\,\FN^\bot = {\rm
    dim}\,\FM^\bot=\frac{1}{2}{\rm dim}\,\mathbb{H}'
=m,
\end{eqnarray}
which means that ${\rm dim}\,\mathbb{H}'$ should be even for
(\ref{generic}) to hold \footnote{Eq.~(\ref{eq:22}) can be derived by
  noting that $\FM^\bot\cap\FN=0$ implies ${\rm ker}(\Q'\P')=0$.
  Indeed, if $Q|p\rangle=0$, where $|p\rangle\in\FN$, then
  $\FM^\bot\cap\FN\not=0$. Hence $|p\rangle=0$. Let us mention for
  completeness that ${\rm ran}(\Q'\P')\cap \FM^\bot =0$.  Indeed, let
  us assume that $|f\rangle\in\FM$, $|g\rangle\in\FN$ and $\langle
  f|\Q' g\rangle=0$. Then $\langle \Q' f|g\rangle=\langle f|
  g\rangle=0$. The last relation means that either $f=0$, or
  $\FM^\bot\cap\FN\not =0$, which contradicts to (\ref{generic}).}.

Here is the statement of the CS-representation \cite{halmos}: after a
unitary transformation $\Q'$ and $\P'$ can be presented as
\begin{eqnarray}
  \label{eq:cscs}
\Q'=\left(
     \begin{array}{cc}
\I_m & \0_m \\
\0_m & \0_m
\end{array}\right), ~~~
\P'=\left(
     \begin{array}{cc}
C^2 & CS \\
CS & S^2
\end{array}\right), ~~~ C^2+S^2=\I_m,~~~ {\rm ker}[C]={\rm ker}[S]=0,
\end{eqnarray}
where all blocks in (\ref{eq:cscs}) have the same dimension $m$.

To prove (\ref{eq:cscs}), note that $\Q'$ and $\P'$ can be
written as [cf. (\ref{eq:22})]
\begin{eqnarray}
  \label{hu}
\Q'=\left(
     \begin{array}{cc}
\I_m & \0_m \\
\0_m & \0_m
\end{array}\right), ~~~
\P'=\left(
     \begin{array}{cc}
K' & B \\
B^\dagger & L
\end{array}\right), ~~~K'\geq 0,~L\geq 0,~~~ {\rm tr}(K'+L)=m.
\end{eqnarray}

Next, let us show that 
\begin{eqnarray}
  \label{eq:74}
  {\rm ker}[\,B \,]=0 
\end{eqnarray}
Since $\Q' \P'(\I-\Q') = \left(
     \begin{array}{cc}
\0_m & B \\
\0_m & \0_m
\end{array}\right)$, we need to show that for any $|\psi\rangle\in
\FM^\bot$, $\Q'\P'|\psi\rangle=0$ means $|\psi\rangle=0$. Indeed, we
have $\P'|\psi\rangle\in \FM^\bot$, which together with
$\FM^\bot\cap\FN=0$ [see (\ref{generic})] leads to $|\psi\rangle=0$.

Eq.~(\ref{eq:74}) implies that there is the well-defined polar
decomposition [$\hat{B}$ is hermitean, while $V$ is unitary]
\begin{eqnarray}
  \label{eq:65}
 B=V\hat{B}, ~~ \hat{B}=\sqrt{B^\dagger B} =\hat{B}^\dagger, ~~
V=B(\,B^\dagger B\,)^{-1/2}=V^{\dagger\,\, -1}.
\end{eqnarray}
We transform as
\begin{eqnarray}
  \label{eq:14}
  \left(
     \begin{array}{cc}
V^\dagger & \0_m \\
\0_m & \I_m
\end{array}\right)\Q'  \left(
     \begin{array}{cc}
V & \0_m \\
\0_m & \I_m
\end{array}\right)=\Q', ~~
  \left(
     \begin{array}{cc}
V^\dagger & \0_m \\
\0_m & \I_m
\end{array}\right)\P'  \left(
     \begin{array}{cc}
V & \0_m \\
\0_m & \I_m
\end{array}\right)=
\left(
     \begin{array}{cc}
K & \hat{B} \\
\hat{B} & L
\end{array}\right),
\end{eqnarray}
where $K=V^\dagger V' U$. We shall now employ the fact that the last
matrix in (\ref{eq:14}) is a projector:
\begin{eqnarray}
  \label{eq:23}
  K=K^2+\hat{B}^2, ~~ L=L^2+\hat{B}^2, ~~ \hat{B}=\hat{B}K+L\hat{B}.
\end{eqnarray}
The first and second relations in (\ref{eq:23}) show that
$[K,\hat{B}]=[L,\hat{B}]=0$. Then the third relations produces
$\hat{B}(K+L-1)=0$. Since $\hat{B}>0$ (due to ${\rm ker} (B)=0$), we
conclude that $K+L=1$. The rest is obvious. 

\subsection{3.2 Joint commutant for two projectors}

Given (\ref{eq:cscs}), we want to find matrices that commute both with
$\P'$ and $\Q'$ \cite{halmos}. Matrices that commute with $\Q'$ read
\begin{eqnarray}
  \label{eq:30}
  \left(
     \begin{array}{cc}
X & \0_m\\
\0_m & Y
\end{array}\right).
\end{eqnarray}
Employing (\ref{eq:cscs}), we get that (\ref{eq:30}) commutes with
$\P'$ if
\begin{eqnarray}
  \label{eq:42a}
&&  XC^2 = C^2X, \\ 
  \label{eq:42b}
&&  YS^2 = S^2Y, \\
  \label{eq:42c}
&&  XCS  = CSY.
\end{eqnarray}
Since $C$ and $S$ are invertible, (\ref{eq:42a}, \ref{eq:42b}) imply
that $[X,S]=[X,C]=[Y,S]=[Y,C]=0$. And then (\ref{eq:42c}) implies that
$X=Y$. Hence 
\begin{eqnarray}
  \label{eq:31}
  X=Y, ~~~~ [X,C]=[X,S]=0.
\end{eqnarray}

\subsection{3.3 General form of the CS representation}

The above derivation of (\ref{eq:cscs}) assumed conditions
(\ref{generic}). More generally, the Hilbert space $\mathbb{H}$ can be
represented as a direct sum \cite{dix,halmos,hardegree}
\begin{eqnarray}
  \label{a:36}
\mathbb{H}=\mathbb{H}'\oplus \mathbb{H}_{11}
\oplus \mathbb{H}_{10}\oplus \mathbb{H}_{01}\oplus
\mathbb{H}_{00},
\end{eqnarray}
where the sub-space $\mathbb{H}_{\alpha\beta}$ of dimension
$m_{\alpha\beta}$ is formed by common eigenvectors of $\p$ and $\q$
having eigenvalue $\alpha$ (for $\p$) and $\beta$ (for
$\q$). Depending on $\p$ and $\q$ every sub-space can be absent;
all of them can be present only for ${\rm dim}\mathbb{H}\geq 6$. Now
$\mathbb{H}_{11}={\rm ran}(\p)\cap{\rm ran}(\q)$.
 $\mathbb{H}'$ has even dimension $2m$
\cite{halmos,hardegree}, this is the only sub-space that is not formed
by common eigenvectors of $\p$ and $\q$.

After a unitary transformation 
\begin{eqnarray}
  \label{eq:75}
\p=U\P U^\dagger, ~~~ \q=U\Q U^\dagger, ~~~ UU^\dagger=\I,
\end{eqnarray}
$\P$ and $\Q$ get the following block-diagonal form that is related to
(\ref{a:36}) \cite{halmos} and that generalizes (\ref{eq:cscs}):
\begin{eqnarray}
  \label{a:cs}
  \Q=Q'\oplus \i_{m_{11}} \oplus \i_{m_{10}} \oplus \0_{m_{01}}\oplus
  \0_{m_{00}}, ~~ Q'\equiv\left(
     \begin{array}{cc}
\i_m & \0_m \\
\0_m & \0_m \\
\end{array}\right),  \\
  \P=P'\oplus \i_{m_{11}} \oplus \0_{m_{10}} \oplus \i_{m_{01}} \oplus \0_{m_{00}}, ~~
P'\equiv\left(
     \begin{array}{cc}
C^2 & CS \\
CS & S^2 \\
\end{array}\right), 
  \label{a:cs1}
\end{eqnarray}
where $C$ and $S$ are invertible square matrices of the same size
holding
\begin{eqnarray}
  \label{a:35}
C^2+S^2=\i_m, ~~ [C,S]=0. 
\end{eqnarray}
$\mathbb{H}'$ refers to $P'$ and $Q'$. If $\P$ and $\Q$ do not have
any common eigenvector, $\P=P'$ and $\Q=Q'$.

\section{4. Derivation of the main result (Eqs.~(\ref{eq:33} ,
  \ref{eq:333}) of the main text)}  

We start with representation (\ref{a:cs}, \ref{a:cs1}) and axioms
(\ref{ax1}--\ref{ax5}) of the main text. These axioms hold for $\P$,
$\Q$ and $\hat{\rho}=U\rho U^\dagger$ [see (\ref{eq:75})] instead of
$\p$, $\q$ and $\rho$, because $\omega(\P,\Q)=U^\dagger\omega(\p,\q)U$
for $\omega=\lo,\uo$ (recall that $\lo(\P,\Q)$ and $\uo(\P,\Q)$ are
Taylor expandable). Hence we now search for $\lo(\P,\Q)$ and
$\uo(\P,\Q)$.

The block-diagonal form (\ref{a:cs}, \ref{a:36}) remains intact
under addition and multiplication of $\P$ and $\Q$. Hence $\lo(\P,\Q)$ and
$\uo(\P,\Q)$ have the block-diagonal form similar to (\ref{a:cs}),
where the diagonal blocks are to be determined. 
Let now $\Pi_{\alpha\beta}$ be the projector on
$\mathbb{H}_{\alpha\beta}$. We get [$\alpha,\beta=0,1$]
\begin{eqnarray}
  \label{a:40}
\Pi_{\alpha\beta}\omega(\P,\Q)\Pi_{\alpha\beta}
=\omega(\alpha,\beta)\Pi_{\alpha\beta},  ~~ \omega=\lo,\,\uo.
\end{eqnarray}
Hence condition (\ref{ax3}) of the main text implies [for
$\omega=\lo,\,\uo$ and $\omega'=\lo',\uo'$]
\begin{eqnarray}
  \label{a:dobr}
  \omega(\P,\Q)=\omega'\oplus \i_{m_{11}} \oplus \0_{m_{10}+m_{01}+m_{00}}, ~~
\omega'=\left(
     \begin{array}{cc}
       \omega'_{11} & \omega'_{12} \\
       {{\omega}'_{12}}^{\dagger} & \omega'_{22} \\
\end{array}\right). 
\end{eqnarray}
Aiming to apply (\ref{ax4}) of the main text, 
we write down (\ref{a:cs}) explicitly as 
\begin{eqnarray}
  \label{a:21}
\Q =
\left(
     \begin{array}{cccccc}
I & 0 & 0 & 0 & 0 & 0\\
0 & 0 & 0 & 0& 0 & 0\\
0 & 0 & I & 0& 0 & 0\\
0 & 0 & 0 & I& 0 & 0\\
0 & 0 & 0 & 0& 0 & 0\\
0 & 0 & 0 & 0& 0 & 0\\
\end{array}\right).
\end{eqnarray}
The most general density matrix $\hat{\rho}$ that commutes with $\Q$
reads (in the same block-diagonal form)
\begin{eqnarray}
  \label{a:32}
\hat{ \rho} =
\left(
     \begin{array}{cccccc}
a_{11} & 0 & a_{12} & a_{13}& 0 & 0\\
0 & b_{11} & 0 & 0& b_{12} & b_{13}\\
a_{12}^\dagger & 0 & a_{22} & a_{23}& 0 & 0\\
a_{13}^\dagger & 0 & a_{23}^\dagger & a_{33}& 0 & 0\\
0 & b_{12}^\dagger & 0 & 0& b_{22} & b_{23}\\
0 & b_{13}^\dagger & 0 & 0& b_{23}^\dagger & b_{33}\\
\end{array}\right).
\end{eqnarray}
Now $\hat{\rho}\Q=\Q\hat{\rho}$ is seen from the fact that 
after permutations of rows and columns, $\Q$ and $\hat{ \rho}$
become $\i_{m+m_{11}+m_{10}}\oplus \0_{m+m_{01}+m_{00}}$ and $a\oplus
b$, respectively. Note that $a_{ii}\geq 0$ and $b_{ii}\geq 0$.

Eqs.~(\ref{a:dobr}, \ref{a:21}, \ref{a:32}) imply
\begin{eqnarray}
  \label{a:34}
  && {\rm tr}(\Q\hat{\rho} \P )={\rm tr}(\Q\hat{\rho} \Q\P )
={\rm tr}(C^2 a_{11}+a_{22}), \\
\label{a:344}
&& {\rm tr}(\lo(\P,\Q)\hat{\rho} )={\rm tr}(\lo'_{11} a_{11} +\lo'_{22}a_{44}+
a_{22}), \\
&& {\rm tr}(\uo(\P,\Q)\hat{\rho} )={\rm tr}(\uo'_{11} a_{11} +\uo'_{22}a_{44}+
a_{22}). 
\label{a:3444}
\end{eqnarray}
Condition (\ref{ax5}) [of the main text] and (\ref{eq:31}) 
imply
\begin{eqnarray}
  \label{eq:78}
\omega'_{11}=\omega'_{22},~~~ \omega'_{12}=0, ~~~ {\rm for}~~~
\omega'=\uo', \lo'.  
\end{eqnarray}
Recall condition (\ref{ax4}) of the main text. It amounts to
(\ref{a:34}) $\geq$ (\ref{a:344}) that should hold for arbitrary $a_{ik}$
and $b_{ik}$. Hence we deduce: $\lo'_{22}=0$ and hence
$\lo'_{11}=\lo'_{12}=0$.  Likewise, (\ref{a:3444}) $\geq$ (\ref{a:34})
leads to $\uo'_{11}=\uo'_{22}=C^2$, $\uo'_{12}=0$; recall that we want
the smallest upper probability. Now (\ref{ax2}) [of the main text]
holds, since
\begin{eqnarray}
  \label{a:39}
\left(
     \begin{array}{cc}
C^2 & 0_m \\
0_m & C^2 \\
\end{array}\right)=\left(
     \begin{array}{cc}
\i_m & 0_m \\
0_m & \i_m \\
\end{array}\right)-(P'-Q')^2. 
\end{eqnarray}
Thus [cf. (\ref{a:36})]
\begin{eqnarray}
  \label{a:37}
&&\uo(\P,\Q)= 
\left(
     \begin{array}{cc}
C^2 & 0_m \\
0_m & C^2 \\
\end{array}\right)\oplus \i_{m_{11}}\oplus \0_{m_{10}+m_{01}+m_{00}}, \\
&&\lo(\P,\Q)= 
\0_{2m}\oplus \i_{m_{11}}\oplus \0_{m_{10}+m_{01}+m_{00}}.
  \label{a:3737}
\end{eqnarray}

Now $\g(\P,\Q)=\g(\Q,\P)$ is the projector onto ${\rm
  ran}(\P)\cap{\rm ran}(\Q)$. To return from (\ref{a:37},
\ref{a:3737}) to original projectors $\p$ and $\q$, we note via
(\ref{a:cs}, \ref{a:cs1}) [recall that $\Pi_{\alpha\beta}$ is the
projector onto $\mathbb{H}_{\alpha\beta}$]:
\begin{eqnarray}
  \label{a:73}
  \Pi_{11}&=&g(\P,\Q), ~~   \Pi_{10}=g(\P,\I-\Q), ~~
  \Pi_{01}=g(\I-\P,\Q), ~~    \Pi_{00}=g(\I-\P,\I-\Q),\\
 \lo(\P,\Q)&=&g(\P,\Q), \\
 P'&=&\P-g(\P,\Q)-g(\P,\I-\Q), \\ 
 Q'&=&\Q-g(\P,\Q)-g(\Q,\I-\P),\\
\uo(\P,\Q)&=&\I-g(\P,\I-\Q)-g(\I-\P,\Q)-g(\I-\P,\I-\Q)\nonumber\\
&& -\left(\, P-Q-g(\P,\I-\Q)+g(\I-\P,\Q)\,\right)^2.
\end{eqnarray}
We act back by $U$, e.g. $g(\P,\Q)=U^\dagger
g(\p,\q)U$, and get finally
\begin{eqnarray}
  \label{a:33}
&&  \lo(\p,\q)=  \g(\p,\q),\\
&& \uo (\p,\q)= \I-(\p-\q)^2-\g(\I-\p,\I-\q).
  \label{a:333}
\end{eqnarray}

\section{5. Representations of upper and lower probability operators}

\subsection{5.1 Representations for the upper probability operator}

Let us turn into a more detailed investigation of (\ref{a:333}). Note
from (\ref{a:cs}, \ref{a:cs1}, \ref{a:73}) that $\I-\g(\I-\p,\I-\q)$
is the projector to ${\rm ran}(\p)+{\rm ran}(\q)$, where ${\rm
  ran}(\p)+{\rm ran}(\q)$ is the vector sum of two sub-spaces. Note
the following representation \cite{piziak}:
\begin{eqnarray}
  \label{eq:71}
  \I-\g(\I-\p,\I-\q)=(\p+\q)^-(\p+\q)=(\p+\q)(\p+\q)^-,
\end{eqnarray}
where $A^-$ is the pseudo-inverse of hermitean $A$, i.e. if $A=V(\,
a\oplus 0\,)V^\dagger$ (where $V$ is unitary: $VV^\dagger=\I$), then
$A^-=V(\, a^{-1}\oplus 0\,)V^\dagger$. 

The third equality in (\ref{eq:71}) is the obvious feature of the
pseudo-inverse. The first equality in (\ref{eq:71}) follows from the
fact that $(\p+\q)^-(\p+\q)$ is the projector on ${\rm ran}(\p+\q)$
and the known relation \cite{piziak}:
\begin{eqnarray}
  \label{eq:66}
  {\rm ran}(\p+\q)={\rm ran}(\p)+{\rm ran}(\q).
\end{eqnarray}

Employing $(\p-\q)^2=\I-(\I-\p-\q)^2$, $\uo (\p,\q)$ can be presented
as a function of $\p+\q$ [cf. (\ref{eq:71})]:
\begin{eqnarray}
\uo (\p,\q) = (\I-\p-\q)^2-\I+(\p+\q)(\p+\q)^-.
\end{eqnarray}

Note another representation for the
projector to ${\rm ran}(\p)+{\rm ran}(\q)$ \cite{jauch_book}
\begin{eqnarray}
  \label{eq:39}
\I-\g(\I-\p,\I-\q)= {\rm min}[A\,|\, A\geq \q, \p],
\end{eqnarray}
where $\I-\g(\I-\p,\I-\q)$ equals to the minimal hermitean operator
$A$ that holds 2 conditions after $|$.

\subsection{5.2 Representations for the lower probability operator}

Let us first show that the projector $g(\p,\q)$ into ${\rm
  ran}(\p)\cap{\rm ran}(\q)$ holds \cite{anderson_duffin}
\begin{eqnarray}
  \label{eq:44}
g(\p,\q)=2\p (\p+\q)^- \q= 2\q (\p+\q)^- \p, 
\end{eqnarray}
where $A^-$ is the pseudo-inverse of $A$ [cf. (\ref{eq:71})]. 

The last equality in (\ref{eq:44}) follows from the fact that $(\q+\p)
(\p+\q)^- $ is the projector to ${\rm ran}(\p)+{\rm ran}(\q)$ [see
(\ref{eq:71}, \ref{eq:66})], which then leads to $\p(\q+\p) (\p+\q)^-
=(\q+\p) (\p+\q)^- \,\p$.

The first equality in (\ref{eq:44}) is shown as follows. Let
$|\psi\rangle\in {\rm ran}(\p)\cap {\rm ran}(\q)$. Then 
using (\ref{eq:44}):
$$2\p (\p+\q)^-
\q |\psi\rangle=(\p+\q)(\p+\q)^- |\psi\rangle=|\psi\rangle.$$ Thus,
${\rm ran}(2\p (\p+\q)^- \q)\supseteq (\, {\rm ran}(\p)\cap {\rm
  ran}(\q)\,)$. On the other hand, ${\rm ran}(2\p (\p+\q)^-
\q)\subseteq {\rm ran}(\p)$ and ${\rm ran}(2\p (\p+\q)^- \q)\subseteq
{\rm ran}(\q)$, where the first relation follows from the implication:
if $|\psi\rangle\not\in {\rm ran}(\p)$, then $|\psi\rangle\not\in {\rm
  ran}(2\p (\p+\q)^- \q)$.

There are two other (more familiar) representations of 
$g(\p,\q)$ [see e.g. \cite{jauch_book,piziak}]:
\begin{eqnarray}
  \label{eq:38}
  \g(\p,\q)&=&{\rm lim}_{n\to\infty}\q(\p\q)^n, \\
           &=& {\rm max}[A\,|\,0\leq A\leq \q, \p].
  \label{gora}
\end{eqnarray}
Eq.~(\ref{eq:38}) can be interpreted as a result of (infinitely many)
successive measurements of $\p$ and $\q$. Eq.~(\ref{gora}) should be
compared to (\ref{eq:39}). 

Yet another representation is useful in calculations, since it
explicitly involves a $2\times 2$ block-diagonal representation
\cite{anderson_duffin}:
\begin{eqnarray}
  \label{eq:79}
&& \p=\left(
     \begin{array}{cc}
\p_{11} & \p_{12} \\
\p_{21} & \p_{22} \\
\end{array}\right), ~~~ \q=\left(
     \begin{array}{cc}
\I_{n_1} & 0 \\
0 & 0_{n_2} \\
\end{array}\right), \\
&& g(\p,\q)=\left(
     \begin{array}{cc}
\p_{11}-\p_{12}\p_{22}^{-}\p_{21} & 0 \\
0 & 0 \\
\end{array}\right),
\label{krot}
\end{eqnarray}
where $\p_{11}$, $\p_{12}$, $\p_{21}$ and $\p_{22}$ are, respectively,
$n_1\times n_1$, $n_1\times n_2$, $n_2\times n_1$, $n_2\times n_2$
matrices.

\subsection{5.3 Direct relation between the eigenvalues of $\p-\q$ and
  $\p\q$}

We can now prove directly (i.e. without employing the CS
representation) that there is a direct relation between the
eigenvalues of $\uo(\p,\q)$ and $\p\q$.  Let $|x\rangle$ be the
eigenvector of hermitean operator $\p-\q$:
\begin{eqnarray}
  \label{eq:72}
  (\p-\q)|x\rangle=\lambda |x\rangle,
\end{eqnarray}
where $-1\leq\lambda\leq 1$ is the eigenvalue. Multiplying both sides
of (\ref{eq:72}) by $\p$ (by $\q$) and using $\p^2=\p$ ($\q^2=\q$) we
get
\begin{eqnarray}
  \label{eq:73}
  \q\p|x\rangle=(1+\lambda)\q |x\rangle, ~~~
  \p\q|x\rangle=(1-\lambda)\p |x\rangle,
\end{eqnarray}
which then implies
\begin{eqnarray}
  \label{eq:733}
  \p\q\,(\p|x\rangle)=(1-\lambda^2)\,(\p |x\rangle), ~~~
  \q\p\,(\q|x\rangle)=(1-\lambda^2)\,(\q |x\rangle).
\end{eqnarray}
Thus $\p|x\rangle$ ($\q|x\rangle$) is an eigenvector of $\p\q$
($\q\p$) with eigenvalue $1-\lambda^2$. 

As seen from (\ref{eq:73}), the 2d linear space ${\rm
  Span}(\p|x\rangle, \q|x\rangle)$ formed by all superpositions of
$\p|x\rangle$ and $\q|x\rangle$ remains invariant under action of both
$\P$ and $\Q$. Together with ${\rm tr}(\p-\q)=0$ this means that if
(\ref{eq:72}) holds, then $\p-\q$ has eigenvalue $-\lambda$ with the
eigen-vector living in ${\rm Span}(\p|x\rangle, \q|x\rangle)$.

Further details on the relation between $\p\q$ and $\p-\q$ can be
looked up in \cite{trapp}.

\section{6. Additivity and monotonicity}

We discuss here the behavior of $\lo(\p,\q)$ and $\uo(\p,\q)$ [given
by (\ref{a:33}, \ref{a:333})] with respect to a monotonic change of
their arguments. For two projectors $\q'$ and $\q$, $\q'\geq\q$ means
$\q'=\q+\k$, where $\k^2=\k$ and $\q\k=0$. Now (\ref{a:33},
\ref{a:333}) and (\ref{gora}) imply that $\lo(\p,\q)$ is operator
superadditive
\begin{eqnarray}
  \label{eq:41a}
  \lo(\p,\q+\k)\geq   \lo(\p,\k)+  \lo(\p,\q).
\end{eqnarray}
Likewise, $\uo(\p,\q)$ is operator subadditive, but
under an additional condition:
\begin{eqnarray}
\uo(\p,\q+\k)\leq   \uo(\p,\k)+  \uo(\p,\q),~~ {\rm if}
~~  \q+\k=\I.
  \label{eq:411a}
\end{eqnarray}
They are the
analogues of classical features (\ref{eq:4}) and (\ref{eq:5}),
respectively. Note that (\ref{eq:41a}) and (\ref{eq:4}) are valid under
the same conditions, since $ \q \k=0$ is the analogue of $A\cap
B=\{0\}$. In that sense the correspondence between (\ref{eq:411a}) and
(\ref{eq:5}) is more limited, since $\q+\k=\I$ is more
restrictive than $\q \k=0$.

We focus on deriving (\ref{eq:41a}), since (\ref{eq:411a}) is derived
in the same way. Note from (\ref{gora}) that $\g(\p,\q)\leq \q$ and
$\g(\p,\k)\leq \k$ imply $\g(\p,\q)+\g(\p,\k)\leq \q+\k$. Since
$\q\k=0$, $g(\p,\q)+g(\p,\k)\leq \p$.  Using (\ref{gora}) for
$g(\p,\q+\k)$ we obtain (\ref{eq:41a}).

Note as well that both $\uo(\p,\q)$ and $\lo(\p,\q)$ are monotonous
[$\omega=\lo,\uo$]:
\begin{eqnarray}
  \label{eq:68}
 \omega (\p',\q') \geq  \omega (\p,\q), ~~{\rm if}~~ \p'\geq \p,\q ~~
 {\rm and}~~
\q'\geq \p,\q.
\end{eqnarray}
Eq.~(\ref{eq:68}) for $\omega=\lo$ follows from (\ref{eq:41a}). For
$\omega=\uo$ it is deduced as follows [cf. (\ref{gora}, \ref{eq:39})]:
\begin{eqnarray}
  \label{eq:69}
  \uo(\p',\q')\geq   \lo(\p',\q')=g(\p',\q')\geq 
  I-g(I-\p,I-\q)\geq 
  \uo(\p,\q).
\end{eqnarray}

Let us now discuss whether (\ref{eq:411a}) can hold under the same
condition $ \q \k=0$ as (\ref{eq:41a}). Now
\begin{eqnarray}
  \label{eq:49}
  \uo(\p,\q+\k)\leq\uo(\p,\q)+\uo(\p,\k),
\end{eqnarray}
amounts to 
\begin{eqnarray}
  \label{eq:50}
  g(\I-\p,\I-\q)+g(\I-\p,\I-\k)\leq \I-\p+g(\I-\p,\I-\q-\k).
\end{eqnarray}
First of all note that for $\q+\k=1$ and $\q\k=0$ we get
$(\I-\q)(\I-\k)=0$ and (\ref{eq:50}) does hold for the same reason as
(\ref{eq:41a}).

For $\q\k=0$, (\ref{eq:50}) is invalid in 3d space (as well as for
larger dimensional Hilbert spaces). Indeed, let us assume that $\q$
and $\k$ are 1d:
\begin{eqnarray}
  \label{eq:51}
\q=\left(
     \begin{array}{ccc}
1 & 0 & 0 \\
0 & 0 & 0 \\
0 & 0 & 0 \\
\end{array}\right), ~~ \k=\left(
     \begin{array}{ccc}
0 & 0 & 0 \\
0 & 0 & 0 \\
0 & 0 & 1 \\
\end{array}\right)  
\end{eqnarray}
Given $\I-\p$ as
\begin{eqnarray}
  \label{eq:52}
\I-\p=\left(
     \begin{array}{ccc}
a_{11} & a_{12} & a_{13} \\
a_{21} & a_{22} & a_{23} \\
a_{31} & a_{32} & a_{33} \\
\end{array}\right)  =
\left(
     \begin{array}{ccc}
a_{11} & a_{12} & a_{13} \\
a_{12}^* & a_{22} & a_{23} \\
a_{13}^* & a_{23}^* & a_{33} \\
\end{array}\right),
\end{eqnarray}
we get [cf. (\ref{krot})]
\begin{eqnarray}
  \label{eq:53}
  g(\I-\p,1-\k)= \left(
     \begin{array}{ccc}
a'_{11} & a'_{12} & 0 \\
a'_{21} & a'_{22} & 0 \\
0 & 0 &0 \\
\end{array}\right), ~~ \left(
     \begin{array}{cc}
a'_{11} & a'_{12}  \\
a'_{21} & a'_{22} \\
\end{array}\right)=\left(
     \begin{array}{cc}
a_{11} & a_{12}  \\
a_{21} & a_{22} \\
\end{array}\right)-\left(
     \begin{array}{c}
a_{13} \\
a_{23} \\
\end{array}\right)\, a_{33}^- \,(a_{31}\, a_{32}), 
\nonumber
\end{eqnarray}
where $a_{33}^-$ is the pseudo-inverse of $a_{33}$.

Likewise,
\begin{eqnarray}
  \label{eq:530}
  g(\I-\p,\I-\q)= \left(
     \begin{array}{ccc}
0 & 0 & 0 \\
0 & a'_{22} & a'_{23} \\
0 & a'_{32} &a'_{33} \\
\end{array}\right), ~~ \left(
     \begin{array}{cc}
a'_{22} & a'_{23}  \\
a'_{32} & a'_{33} \\
\end{array}\right)=\left(
     \begin{array}{cc}
a_{22} & a_{23}  \\
a_{32} & a_{33} \\
\end{array}\right)-\left(
     \begin{array}{c}
a_{21} \\
a_{31} \\
\end{array}\right)\, a_{11}^- \,(a_{12}\, a_{13}).
\nonumber
\end{eqnarray}
Now $g(\I-\p,\I-\q-\k)=0$, since $\I-\q-\k$ is a 1d projector. We can
now establish that generically 
\begin{eqnarray}
  \label{eq:34}
g(\I-\p,\I-\q)+g(\I-\p,\I-\k)\not\leq \I
\end{eqnarray}
(let alone (\ref{eq:50})), because the
difference has both positive and negative eigenvalues.

The message (\ref{eq:34}) is that the function $\I-\g(\I-\p,\I-\q)$ is
not sub-additive. 

Now consider (\ref{eq:49}, \ref{eq:50}), but under additional
condition that $\p\k=0$. Now (\ref{eq:50}) amounts to 
\begin{eqnarray}
  \label{eq:40}
    g(\I-\p,\I-\q)\leq \k+g(\I-\p,\I-\q-\k),
\end{eqnarray}
which holds as equality since ${\rm ran}(\k) \subseteq {\rm
  ran}(\p^\bot)\cap {\rm ran}(\q^\bot)$.

\section{7. Upper and lower probabilities for simple examples}

\subsection{7.1 Two-dimensional Hilbert space}

It should be clear from (\ref{a:37}, \ref{a:3737}) that in
two-dimensional Hilbert space, any lower probability operator
$\lo(\p,\q)$ is zero (since two rays overlap only at zero), while the
upper probability operator $\uo(\p,\q)=\up(\rho; \p,\q)$ just reduces
to the transition probability (i.e. to a number) ${\rm tr}(\p\q)$.
Thus for the present case both $\up$ and $\lp$ do not depend on
$\rho$. \comment{ For spin $1/2$, the spin components read
\begin{eqnarray}
  \label{eq:47}
\sigma^x=
\left(
     \begin{array}{cc}
0 & 1  \\
1 & 0  \\
\end{array}\right), ~~
\sigma^y=
\left(
     \begin{array}{cc}
0 & -i  \\
i & 0  \\
\end{array}\right), ~~
\sigma^z=
\left(
     \begin{array}{cc}
1 & 0  \\
0 & -1  \\
\end{array}\right).
\end{eqnarray}
For the projectors we have
\begin{eqnarray}
  \label{eq:48}
&&  P^x_{\pm 1} = \frac{1}{2}
\left(
     \begin{array}{cc}
1 & \pm 1  \\
\pm 1 & 1  \\
\end{array}\right), ~~
  P^y_{\pm 1} = \frac{1}{2}
\left(
     \begin{array}{cc}
1 & \mp i  \\
\pm i & 1  \\
\end{array}\right), \\
&&  P^z_1 = 
\left(
     \begin{array}{cc}
1 & 0  \\
0 & 0  \\
\end{array}\right), ~~
 P^z_{-1} = 
\left(
     \begin{array}{cc}
0 & 0  \\
0 & 1  \\
\end{array}\right).
\end{eqnarray}
For non-commuting quantities, all lower probabilities $\lo$ are equal
to zero, while all upper probabilities $\uo$ are equal to
$\frac{1}{2}$. In the two-dimensional situation, $\lo(P,Q)=0$ for
$P\not =Q$, and $\uo(P,Q)\propto I$. 
}

\subsection{7.2 Spin 1}

The $3\times 3$ matrices for the spin components read
\begin{eqnarray}
  \label{eq:300}
  L^x=\frac{1}{\sqrt{2}}\left(
     \begin{array}{ccc}
0 & 1 & 0 \\
1 & 0 & 1 \\
0 & 1 & 0 \\
\end{array}\right), ~~
  L^y=\frac{i}{\sqrt{2}}\left(
     \begin{array}{ccc}
0 & -1 & 0 \\
1 & 0 & -1 \\
0 & 1 & 0 \\
\end{array}\right), ~~
  L^z=\left(
     \begin{array}{ccc}
1 & 0 & 0 \\
0 & 0 & 0 \\
0 & 0 & -1 \\
\end{array}\right).
\end{eqnarray}

Now $P^a_{\pm 1,0}$ for $a=x,y,z$ are the one-dimensional projectors
to the eigenspace with eigenvalues $\pm 1$ or $0$ of $L^a$:
\begin{eqnarray}
  \label{eq:43}
  P^x_{1}=\frac{1}{4}\left(
     \begin{array}{ccc}
1 & \sqrt{2} & 1 \\
\sqrt{2} & 2 & \sqrt{2} \\
1 & \sqrt{2} & 1 \\
\end{array}\right) ~~~ 
  P^x_{0}=\frac{1}{2}\left(
     \begin{array}{ccc}
1 & 0 & -1 \\
0 & 0 & 0 \\
-1 & 0 & 1 \\
\end{array}\right), ~~ 
  P^x_{-1}=\frac{1}{4}\left(
     \begin{array}{ccc}
1 & -\sqrt{2} & 1 \\
-\sqrt{2} & 2 & -\sqrt{2} \\
1 & -\sqrt{2} & 1 \\
\end{array}\right),
\end{eqnarray}
\begin{eqnarray}
  \label{eq:433}
  P^y_{1}=\frac{1}{4}\left(
     \begin{array}{ccc}
1 & -i\sqrt{2} & -1 \\
i\sqrt{2} & 2 & -i\sqrt{2} \\
-1 & i\sqrt{2} & 1 \\
\end{array}\right) ~~~ 
  P^y_{0}=\frac{1}{2}\left(
     \begin{array}{ccc}
1 & 0 & 1 \\
0 & 0 & 0 \\
1 & 0 & 1 \\
\end{array}\right), ~~ 
  P^y_{-1}=\frac{1}{4}\left(
     \begin{array}{ccc}
1 & i\sqrt{2} & -1 \\
-i\sqrt{2} & 2 & i\sqrt{2} \\
-1 & -i\sqrt{2} & 1 \\
\end{array}\right),
\end{eqnarray}
\begin{eqnarray}
  \label{eq:4333}
  P^z_{1}=\left(
     \begin{array}{ccc}
1 & 0 & 0 \\
0 & 0 & 0 \\
0 & 0 & 0 \\
\end{array}\right) ~~~ 
  P^z_{0}=\left(
     \begin{array}{ccc}
0 & 0 & 0 \\
0 & 1 & 0 \\
0 & 0 & 0 \\
\end{array}\right), ~~ 
  P^z_{-1}=\left(
     \begin{array}{ccc}
0 & 0 & 0 \\
0 & 0 & 0 \\
0 & 0 & 1 \\
\end{array}\right),
\end{eqnarray}
where the zero components are orthogonal to each other:
\begin{eqnarray}
  \label{eq:46}
  P^x_{0}  P^y_{0}=  P^x_{0}  P^z_{0} = P^z_{0}  P^y_{0}=0.
\end{eqnarray}
Other overlaps are simple as well $(\alpha\not =\beta)$
\begin{eqnarray}
  {\rm tr}(P^\alpha_j P^\beta_k)&=& 1/4 ~~ {\rm if}~~ j\not =0~{\rm
    and}~ k\not =0,\nonumber\\
  &=& 1/2 ~~ {\rm if}~~ j =0~{\rm
    or}~ k=0 ~ {\rm but~not~both},\nonumber\\
  &=& 0 ~~ {\rm if}~~ j =0~{\rm and}~k=0.
  \label{eq:67}
\end{eqnarray}

Given 2 projectors $P$ and $Q$, we defined $g(P,Q)$ as the projector
on ${\rm ran}(P)\cap {\rm ran}(Q)$.  For calculating $g(P,Q)$ we
employ (\ref{eq:44}). Here are upper probability operators for joint
values of $P^z$ and $P^x$:
\begin{eqnarray}
  \label{gg1}
&&  \uo(P^z_0,P^x_0)=0,\\
  \label{gg2}
&&  \uo(P^x_{\pm 1},P^z_1)=
\left(
     \begin{array}{ccc}
\frac{1}{4} & 0 & 0 \\
0 & \frac{1}{6} & \frac{\pm 1}{6\sqrt{2}} \\
0 & \frac{\pm 1}{6\sqrt{2}} & \frac{1}{12} \\
\end{array}\right),\\
  \label{gg3}
&&  \uo(P^x_{\pm 1},P^z_{-1})=  \left(
     \begin{array}{ccc}
\frac{1}{12} & \frac{\pm 1}{6\sqrt{2}} & 0 \\
\frac{\pm 1}{6\sqrt{2}} & \frac{1}{6} & 0 \\
0 & 0 & \frac{1}{4} \\
\end{array}\right), \\
  \label{gg4}
&&  \uo(P^x_{0},P^z_{\pm 1})=
  \left(
     \begin{array}{ccc}
\frac{1}{2} & 0 & 0 \\
0 & 0 & 0 \\
0 & 0 & \frac{1}{2} \\
\end{array}\right), \\
  \label{gg5}
&& 
\uo(P^x_{\pm 1},P^z_0)=
  \left(
     \begin{array}{ccc}
\frac{1}{4} & 0 & - \frac{1}{4} \\
0 & \frac{1}{2} & 0 \\
-\frac{1}{4} & 0 & \frac{1}{4} \\
\end{array}\right). 
\end{eqnarray}
Since these are 1d projectors, all the lower probability operators
nullify. Now (\ref{gg1}) means that the precise probability of $P^z_0$
and $P^x_0$ is zero; cf. (\ref{eq:46}). 

We now get from (\ref{gg1}--\ref{gg5})
\begin{eqnarray}
  \label{eq:70}
  \sum_{k=\pm, 0}  \sum_{i=\pm, 0}
\uo(P^x_{k},P^z_i)=  \left(
     \begin{array}{ccc}
\frac{13}{6} & 0 & - \frac{1}{2} \\
0 & \frac{5}{3} & 0 \\
-\frac{1}{2} & 0 & \frac{13}{6} 
\end{array}\right). 
\end{eqnarray}
This matrix is larger than $\I$, since its eigenvalues are
$\frac{8}{3}$, $\frac{5}{3}$ and $\frac{5}{3}$.

Note from (\ref{a:cs}, \ref{a:cs1}) that for $3\times 3$ matrices
${\rm dim}{\cal H}'=2$, while ${\rm dim}{\cal H}_{11}=1$ (if this
sub-space is present at all). Hence the eigenvalues of $\uo$ relate to
transition probabilities (\ref{eq:67}). Indeed, the eigenvalues of
matrices in (\ref{gg2}, \ref{gg3}) [resp. in (\ref{gg4}, \ref{gg5})]
is $(\frac{1}{4}, \frac{1}{4},0)$ [resp. $(\frac{1}{2},
\frac{1}{2},0)$]. Hence the maximal probability interval
$[\frac{1}{4},0]$ that can be generated by (\ref{gg2}, \ref{gg3}) is
smaller than the maximal interval $[\frac{1}{2},0]$ generated by
(\ref{gg4}, \ref{gg5}). As an example, let us take the upper
probabilities generated on eigenstates of $L^y$ ($\epsilon, \eta, \chi
=1,0,-1$):
\begin{eqnarray}
  \label{eq:45}
  {\rm tr}[\, \uo(P^x_{\epsilon},P^z_\eta)\, P^y_\chi   \,]  =&& 1/6 ~~
  {\rm if}~~\epsilon\eta\not =0, \nonumber \\
  =&& 1/4 ~~ {\rm if} ~~ \epsilon\eta =0,\, (1-\epsilon)(1-\eta)\not
  =1,\, \chi\not=0, \nonumber \\
  =&& 1/2 ~~ {\rm if} ~~ \epsilon\eta =0,\, (1-\epsilon)(1-\eta)\not
  =1,\, \chi=0, \nonumber \\
  =&& 0 ~~ {\rm if} ~~ \epsilon=\eta=\chi =0.
\end{eqnarray}

Let us now turn to joint probabilities, where the lower probability is
non-zero.
\begin{eqnarray}
\label{kk1}
&&  \lo(P^x_0+P^x_{\pm 1},P^z_1+P^z_{0})=
  \left(
     \begin{array}{ccc}
\frac{2}{3} &\pm\frac{\sqrt{2}}{3} & 0 \\
\pm\frac{\sqrt{2}}{3} & \frac{1}{3} & 0 \\
0 & 0 & 0 \\
\end{array}\right), ~ 
\uo(P^x_0+P^x_{\pm 1},P^z_1+P^z_{0})=
  \left(
     \begin{array}{ccc}
\frac{3}{4} & \pm\frac{1}{2\sqrt{2}} & 0 \\
\pm\frac{1}{2\sqrt{2}} & \frac{1}{2} & 0 \\
0 & 0 & \frac{1}{4} \\
\end{array}\right), \\
\label{kk2}
&&  \lo(P^x_0+P^x_{\pm 1},P^z_{-1}+P^z_{0})=
  \left(
     \begin{array}{ccc}
0 & 0 & 0 \\
0 & \frac{1}{3} & \pm\frac{\sqrt{2}}{3} \\
0 & \pm\frac{\sqrt{2}}{3} & \frac{2}{3} \\
\end{array}\right), ~
\uo(P^x_0+P^x_{\pm 1},P^z_{-1}+P^z_{0})=
\left(
     \begin{array}{ccc}
\frac{1}{4} & 0 & 0 \\
0 & \frac{1}{2} & \pm\frac{1}{2\sqrt{2}} \\
0 & \pm\frac{1}{2\sqrt{2}} & \frac{3}{4} \\
\end{array}\right),\\
\label{kk3}
&&  \lo(P^x_1+P^x_{- 1},P^z_{\pm 1}+P^z_{0})=
  \left(
     \begin{array}{ccc}
0 & 0 & 0 \\
0 & 1 & 0 \\
0 & 0 & 0 \\
\end{array}\right), ~
 \uo(P^x_1+P^x_{- 1},P^z_{\pm 1}+P^z_{0})=
  \left(
     \begin{array}{ccc}
\frac{1}{2} & 0 & 0 \\
0 & 1 & 0 \\
0 & 0 & \frac{1}{2} \\
\end{array}\right), \\
\label{kk4}
&&  \lo(P^x_0+P^x_{\pm 1},P^z_{1}+P^z_{-1})=
  \left(
     \begin{array}{ccc}
\frac{1}{2} & 0 & -\frac{1}{2} \\
0 & 0 & 0 \\
-\frac{1}{2} & 0 & \frac{1}{2} \\
\end{array}\right), ~
\uo(P^x_0+P^x_{\pm 1},P^z_{1}+P^z_{-1})=
\left(
     \begin{array}{ccc}
\frac{3}{4} & 0 & -\frac{1}{4} \\
0 & \frac{1}{2} & 0 \\
-\frac{1}{4} & 0 & \frac{3}{4} \\
\end{array}\right), \\
\label{kk5}
&&  \lo(P^x_1+P^x_{- 1},P^z_{1}+P^z_{-1})=\uo(P^x_1+P^x_{- 1},P^z_{1}+P^z_{-1})=
  \left(
     \begin{array}{ccc}
\frac{1}{2} & 0 & \frac{1}{2} \\
0 & 0 & 0 \\
\frac{1}{2} & 0 & \frac{1}{2} \\
\end{array}\right).
\end{eqnarray}
Now $\uo-\lo$ for (\ref{kk1}, \ref{kk2}) has eigenvalues
$(\frac{1}{4}, \frac{1}{4},0)$, while for for (\ref{kk3}, \ref{kk4})
this matrix has eigenvalues $(\frac{1}{2}, \frac{1}{2},0)$ (the last
case (\ref{kk5}) refers to the commutative situation). Hence the
probabilities for (\ref{kk3}, \ref{kk4}) are more uncertain.

Next, let us establish whether certain combinations can be (surely)
more probable than others. Note that
\begin{eqnarray}
  \label{eq:56}
{\rm Eigenvalues}[\,\lo(P^x_0+P^x_{-
  1},P^z_{1}+P^z_{0})-\uo(P^x_0+P^x_{1},P^z_{1}
+P^z_{0})\,]=\left(
\frac{\pm\sqrt{393}-3}{24},\, -\frac{1}{4}
\right).  
\end{eqnarray}
Once there is (one) positive eigenvalue, there is a class of states $\rho$
for which 
\begin{eqnarray}
  \label{eq:54}
{\rm tr}(\,\rho\, \lo(P^x_0+P^x_{-
  1},P^z_{1}+P^z_{0})  \,)>  {\rm tr}(\,\rho\, \uo(P^x_0+P^x_{
  1},P^z_{1}+P^z_{0})  \,),  
\end{eqnarray}
i.e. $P^x=0~{\rm or}~-1$ and $P^z=0~{\rm or}~1$ is more probable than 
$P^x=0~{\rm or}~1$ and $P^z=0~{\rm or}~1$. Note that 
\begin{eqnarray}
  \label{eq:58}
  [\,\lo(P^x_0+P^x_{-
  1},P^z_{1}+P^z_{0}) \,,\,\uo(P^x_0+P^x_{
  1},P^z_{1}+P^z_{0})\,]\not =0.
\end{eqnarray}
Such examples can be easily
continued, e.g.
\begin{eqnarray}
  \label{eq:57}
{\rm Eigenvalues}[ \,\lo(P^x_0+P^x_{
  1},P^z_{1}+P^z_{0})-\uo(P^x_{1}+P^x_{-1},P^z_{1}
+P^z_{0})\,]=\left(
\frac{\pm\sqrt{57}-3}{12},\, -\frac{1}{2}
\right).  
\end{eqnarray}

\end{document}